\documentclass[aps,pre,reprint,amsmath,amssymb,floats]{revtex4-2}

\usepackage{graphicx}
\usepackage{bm}
\usepackage{color}
\usepackage{physics}
\usepackage{mathtools}

\begin{document}
\title{Eigenstate thermalization hypothesis and eigenstate-to-eigenstate
fluctuations}
\author{Jae Dong Noh}
\affiliation{Department of Physics, University of Seoul, Seoul 02504, Korea}

\date{\today}

\begin{abstract}
  We investigate the extent to which the eigenstate thermalization
  hypothesis~(ETH) is valid or violated in the nonintegrable and 
  the integrable spin-$1/2$ {\it XXZ} chains. We perform the energy-resolved 
  analysis of statistical properties of matrix elements of observables 
  in the energy eigenstate basis.
  The Hilbert space is divided into energy shells of constant width, and 
  a block submatrix is constructed whose columns and rows correspond to the
  eigenstates in the respective energy shells. In each submatrix, we
  measure the second moment of off-diagonal elements in a column.
  The columnar second moments are distributed with a finite variance for
  finite-sized systems. We show that the relative variance of the 
  columnar second moments decreases as 
  the system size increases in the nonintegrable system.
  The self-averaging behavior indicates that the energy eigenstates are 
  statistically equivalent to each other, which is consistent with the ETH. 
  In contrast, the relative variance does not decrease with the system size 
  in the integrable system. 
  The persisting eigenstate-to-eigenstate fluctuation implies that the
  matrix elements cannot be characterized with the energy parameters only.
  Our result explains the origin for the breakdown of
  the fluctuation dissipation theorem in the integrable system. The
  eigenstate-to-eigenstate fluctuations sheds a new light on the meaning of the ETH.
\end{abstract}

\maketitle

\section{Introduction}\label{sec:intro}

Since the birth of quantum mechanics, it has been a fascinating question to ask whether an isolated quantum system can
approach the thermal equilibrium state through the unitary time evolution~\cite{vonNeumann:1929vj,vonNeumann:2010jl}.
The thermalization requires that the Hamiltonian eigenstate expectation value 
of an observable should be equal to the equilibrium ensemble average and that 
temporal fluctuations and responses should be governed by the fluctuation dissipation 
theorem~(FDT). It is believed that the eigenstate thermalization hypothesis~(ETH)
provides a mechanism for the quantum
thermalization~\cite{Srednicki:1996kn,Rigol:2008bf,DAlessio:2016gr,Deutsch:2018fy}.

An isolated quantum system is characterized by the Hamiltonian $\hat{H}$. 
Let $\{\ket{\alpha} \}$ and $\{ E_\alpha\}$ be the energy eigenstates and the
eigenvalues, respectively. The ETH~\cite{Srednicki:1996kn,Srednicki:1999bo}
proposes that matrix elements 
$O_{\gamma\alpha} \equiv \bra{\gamma} \hat{O} \ket{\alpha}$ of
an observable $\hat{O}$ should take the form of
\begin{equation}
  O_{\gamma\alpha} = O(E_{\gamma\alpha})
  \delta_{\gamma\alpha} + e^{-S(E_{\gamma\alpha})/2} f_O(E_{\gamma\alpha},
  \omega_{\gamma\alpha}) R_{\gamma\alpha} ,
  \label{ETH_ansatz}
\end{equation}
where $E_{\gamma\alpha} \equiv (E_\gamma+E_\alpha)/2$, $\omega_{\gamma\alpha}
\equiv E_\gamma-E_\alpha$, $S(E)$ is the thermodynamic entropy with the Boltzmann
constant $k_B=1$, $R_{\gamma\alpha}$'s are elements of a random matrix in
the Gaussian orthogonal or unitary ensemble, and $O$
and $f_O$ are {\em smooth} functions of their arguments.

According to the ETH, a diagonal element, eigenstate expectation value, is
given by 
\begin{equation}
  O_{\alpha\alpha} = O(E_\alpha) + e^{-S(E_\alpha)/2} f_O(E_\alpha,0)
  R_{\alpha\alpha} .
  \label{ETH_diag}
\end{equation}
Since the entropy is an extensive quantity, the second term 
decreases exponentially with the system size. Thus, the diagonal elements 
follow the Gaussian distribution whose variance is exponentially small. The ETH for off-diagonal elements reads as
\begin{equation}
  O_{\gamma\alpha(\neq\gamma)} = e^{-S(E_{\gamma\alpha})/2}
  f_O(E_{\gamma\alpha},\omega_{\gamma\alpha}) R_{\gamma\alpha}  .
  \label{ETH_offd}
\end{equation}
The off-diagonal elements govern temporal fluctuations. The ETH guarantees
that an isolated quantum system in an energy eigenstate obeys the 
FDT~\cite{Srednicki:1999bo,Khatami:2013kf,DAlessio:2016gr,Noh:2020fs,Schuckert:2020bu}.

The ETH ansatz has been tested extensively for the generic nonintegrable
and the integrable
systems~\cite{Rigol:2009ew,Steinigeweg:2013dc,Ikeda:2013kw,Steinigeweg:2014dj,Beugeling:2014ci,Kim:2014kw,Beugeling:2015hq,Alba:2015hs,Mondaini:2016dn,Mondaini:2017jg,Yoshizawa:2018js,Nation:2018kg,Jansen:2019hu,Noh:2019gx,LeBlond:2019bv,Brenes:2020jx,Richter:2020bx,Schuckert:2020bu}. The former
is known to obey the ETH, while the latter does not obey the
ETH~\footnote{Systems exhibiting many-body localizations or
  scars do not obey the ETH either. They are not discussed in this work.}. 
We present a brief review on the previous numerical works on the ETH.

For the diagonal elements, one may measure the difference 
$\delta O_\alpha = | O_{(\alpha+1)(\alpha+1)}- O_{\alpha\alpha}|$ in the
eigenstate expectation values of neighboring eigenstates~\cite{Kim:2014kw}.
In the nonintegrable systems, both their support and the mean value have
been shown to decrease exponentially with the system
size~\cite{Kim:2014kw,Mondaini:2017jg,Jansen:2019hu,LeBlond:2019bv}. 
These behaviors are consistent with 
Eq.~\eqref{ETH_diag}. On the other hand, in the integral systems, 
they have a nonvanishing support and their mean value decreases 
algebraically or more slowly with the system size~\cite{Alba:2015hs,LeBlond:2019bv}. 

The off-diagonal elements have been shown to follow
the Gaussian distribution in the nonintegrable systems. 
The variance is inversely proportional to the
density of states and decreases exponentially with the system
size~\cite{LeBlond:2019bv,Richter:2020bx}. 
These behaviors are also consistent with the ETH prediction. 
Using the variance and the density of states, one can estimate the function 
$|f_O(E,\omega)|$ numerically~\cite{Mondaini:2017jg}.

The off-diagonal elements have been also investigated in 
the integrable systems. They do not follow the Gaussian
distribution~\cite{Beugeling:2015hq,LeBlond:2019bv}. A model
study~\cite{LeBlond:2019bv} reports that they may follow a log-normal
distribution.
Interestingly, the variance is also found to be inversely proportional 
to the density of
states as in the nonintegrable
systems~\cite{Mallayya:2019kq,LeBlond:2019bv,Brenes:2020jx}.

The off-diagonal elements in the nonintegrable and the integrable systems
share two important features: (i) The variance is inversely proportional to the
density of states, and (ii) it seems to be a smooth function of $E_{\gamma\alpha}$ and
$\omega_{\gamma\alpha}$. These suggest that the off-diagonal elements in
the integrable systems may follow the ansatz of Eq.~\eqref{ETH_offd} 
with non-Gaussian random variables $R_{\gamma\alpha}$. 
We notice that the two common features are the essential ingredients 
leading to the FDT at the energy eigenstates~\cite{DAlessio:2016gr,Noh:2020fs}. 
Apparently, it is 
contradictory to the known fact that the FDT is violated in the integrable
systems~\cite{Foini:2011kw,Foini:2012gq}. 

In this paper, we perform the energy-resolved study of the statistical properties 
of matrix elements $\{O_{\gamma\alpha}\}$ of observables 
in the integrable and in the nonintegrable spin-1/2 {\it XXZ} models. From the full matrix,
one can construct a block submatrix $\tilde{O}$ whose columns and rows
correspond to the energy eigenstates belonging to an energy shell of width
$\Delta_E$. Each block is characterized with constant energy parameters 
$E_{\gamma\alpha}$ and $\omega_{\gamma\alpha}$ up to $\Delta_E$. 
Investigating the probability distribution of the elements within each block,
one can test the ETH ansatz at various energy values. 
Furthermore, one can also investigate an
eigenstate-to-eigenstate fluctuation by comparing the probability
distributions of matrix elements in different columns.

As a main result, we will show that the eigenstate-to-eigenstate fluctuation vanishes in
the nonintegrable system while it remains finite in the integrable system
in the large system size limit. 
The ETH ansatz in Eq.~\eqref{ETH_offd} for the off-diagonal elements requires
that all matrix elements should be equivalent statistically within a block
submatrix for a sufficiently small $\Delta_E$.
Otherwise, the function $f_O(E,\omega)$ cannot be a {\em smooth} function of
the arguments. The eigenstate-to-eigenstate fluctuation 
disproves the existence of such a smooth function in the integrable system.
Thus, it resolves the puzzle in regard to the FDT. 

This paper is organized as follows. In Sec.~\ref{sec:model}, we introduce
the spin-1/2 {\it XXZ} Hamiltonian and the observables investigated in this paper. 
We also explain the method 
to construct block submatrices. In Sec.~\ref{sec:numerical}, we present the
numerical results for the energy-resolved statistics for matrix elements
within each submatrix.
In Sec.~\ref{sec:eef}, we investigate the
eigenstate-to-eigenstate fluctuations in statistics of off-diagonal matrix
elements. It will uncover the clear difference between the integrable system 
and the nonintegrable system. 
We conclude the paper with summary and discussions in Sec.~\ref{sec:summary}.

\section{Model system and numerical setup}\label{sec:model}
We study the spin-1/2 {\it XXZ} model in a one-dimensional lattice of $L$ sites under the
periodic boundary condition. Let $\hat{\sigma}_l^p$ be the Pauli matrix in the
$p~(=x, y, z)$ direction at site $l~(=1,\cdots, L)$. 
The {\it XXZ} Hamiltonian is given by
\begin{equation}
  \hat{H} = \frac{1}{1+\lambda} \sum_{l=1}^L \left[ \hat{h}_{l, l+1} + \lambda
  \hat{h}_{l, l+2}\right] ,
  \label{H_XXZ}
\end{equation}
where $\hat{h}_{l,m} \equiv -\frac{J}{2} (
  \hat{\sigma}_l^x \hat{\sigma}_m^x + \hat{\sigma}_l^y \hat{\sigma}_m^y +
\Delta \hat{\sigma}_l^z \hat{\sigma}_m^z )$. 
Note that $\Delta$ is an anisotropy parameter and $\lambda$ is the relative 
strength of the next nearest neighbor interactions. 
The overall coupling constant $J$ will be set to unity so that the energy 
becomes dimensionless. 
The Hamiltonian is integrable when $\lambda=0$, and
nonintegrable when 
$\lambda\neq 0$~\cite{Yoshizawa:2018js,Noh:2019gx,Noh:2020fs}. 

The Hamiltonian commutes with the total magnetization operator in the $z$ direction,
the translation operator, the spatial reflection operator, and the spin inversion 
operator. Especially, in the translationally invariant subspace with 
zero magnetization, all the symmetry operators mutually 
commute~\cite{LeBlond:2019bv,Jung:2020ia}.  
We focus our attention on the translationally invariant 
subspace with zero magnetization that are even under the spatial reflection and 
the spin inversion,  which will be called the maximum symmetry
sector~(MSS)~\cite{Jung:2020ia}. The Hilbert space dimensions of the MSS
are $D= 2518, 8359$, and $28968$ for $L=20, 22$, and $24$, respectively. 
In this paper, we present the numerical results obtained at
$\lambda=0$~(integrable case) and $\lambda=1$~(nonintegrable case) with fixed 
$\Delta = 1/2$.

We numerically diagonalize the Hamiltonian in the MSS 
to obtain the energy eigenvalues $\{E_\alpha\}$ and the
eigenstates $\{\ket{\alpha}\}$ with $\alpha = 1, \cdots,
D$~\cite{Jung:2020ia,Sandvik:2010kc}. The eigenstates
are arranged in the ascending order of the energy eigenvalues. 
Using the eigenvalue spectrum, we define a function 
\begin{equation}
  \bar{E}(\beta) = \frac{\sum_{\gamma} E_\gamma e^{-\beta E_\gamma}}{\sum_\gamma
    e^{-\beta E_\gamma}} .
  \label{beta_E}
\end{equation}
If the system is thermal, $\bar{E}(\beta)$ corresponds to the equilibrium 
canonical ensemble
average of the energy at the inverse temperature $\beta$.
One can assign the temperature to each energy eigenstate using the relation
$\bar{E}(\beta) = E_\alpha$. 
Such an assignment is useful even in the nonthermal case 
since it allows one to parametrize the energy eigenvalue with an 
intensive variable $\beta$. The parameter $\beta$ will be called the inverse
temperature in both cases for convenience.

\begin{figure}[t]
  \includegraphics*[width=0.9\columnwidth]{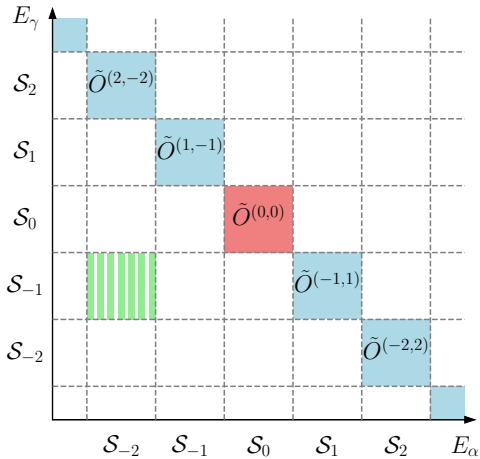}
  \caption{Illustration of the block submatrix structure. Each square
    corresponds to a block submatrix. In Sec.~\ref{sec:numerical}, we study
    the statistical properties of matrix elements in each shaded block. 
    The columnar stripe represents matrix elements involving a given energy
    eigenstate. Statistical fluctuations from column to column, that is, 
    the eigenstate-to-eigenstate fluctuations 
    will be studied in Sec.~\ref{sec:eef}.
  }
  \label{fig:1}
\end{figure}

Once all eigenvectors are obtained, it is straightforward to calculate
the matrix elements $O_{\gamma\alpha}$ of an observable $\hat{O}$.
For a detailed study, we introduce a block submatrix.
To a given value of $\beta$ and $\bar{E} = \bar{E}(\beta)$, 
we define an energy shell $\mathcal{S}_a$~($a=0,\pm 1, \pm2, \cdots$) 
as the subspace consisting of 
energy eigenstates with $\bar{E}+(a-\frac{1}{2}) \Delta_E \le E_\alpha < 
\bar{E}+(a+\frac{1}{2}) \Delta_E$. The energy resolution
$\Delta_E$ is taken to be a constant independent of $L$.
The number of energy eigenstates within a shell $\mathcal{S}_a$ 
is denoted by $\mathcal{N}_a$. 
A block submatrix $\tilde{O}^{(b,a)}(\beta)$ 
is defined as a $(\mathcal{N}_b \times \mathcal{N}_a)$ matrix consisting of elements 
$O_{\gamma\alpha}$ with $\ket{\gamma}\in \mathcal{S}_b$ and
$\ket{\alpha}\in \mathcal{S}_a$. All elements of $\tilde{O}^{(b,a)}$ are
characterized by constant $E_{\gamma\alpha} \simeq \bar{E}(\beta)+(a+b)\Delta_E/2$ and
$\omega_{\gamma\alpha}\simeq (b-a)\Delta_E$ up to $\Delta_E$. We illustrate the block
submatrix structure in Fig.~\ref{fig:1}.

As observables, we choose the nearest neighbor interaction energy
\begin{equation}
  \hat{O}_1 = \frac{1}{\sqrt{L}} \sum_l \hat{\sigma}_l^z \hat{\sigma}_{l+1}^z
  \label{O1_def}
\end{equation}
and the zero momentum distribution function
\begin{equation}
  \hat{O}_2 = \frac{1}{L} \sum_{l,m} \hat{\sigma}_l^+ \hat{\sigma}_m^- .
  \label{O2_def}
\end{equation}
Notice that $\hat{O}_1$ is normalized with $\sqrt{L}$ instead of $L$. With this choice,
the Hilbert-Schmidt norm becomes $L$-independent 
and the ETH analysis becomes easy~\cite{Mierzejewski:2020jm,LeBlond:2019bv}.

\section{Numerical results}\label{sec:numerical}
\begin{figure*}[t]
  \includegraphics*[width=2.0\columnwidth]{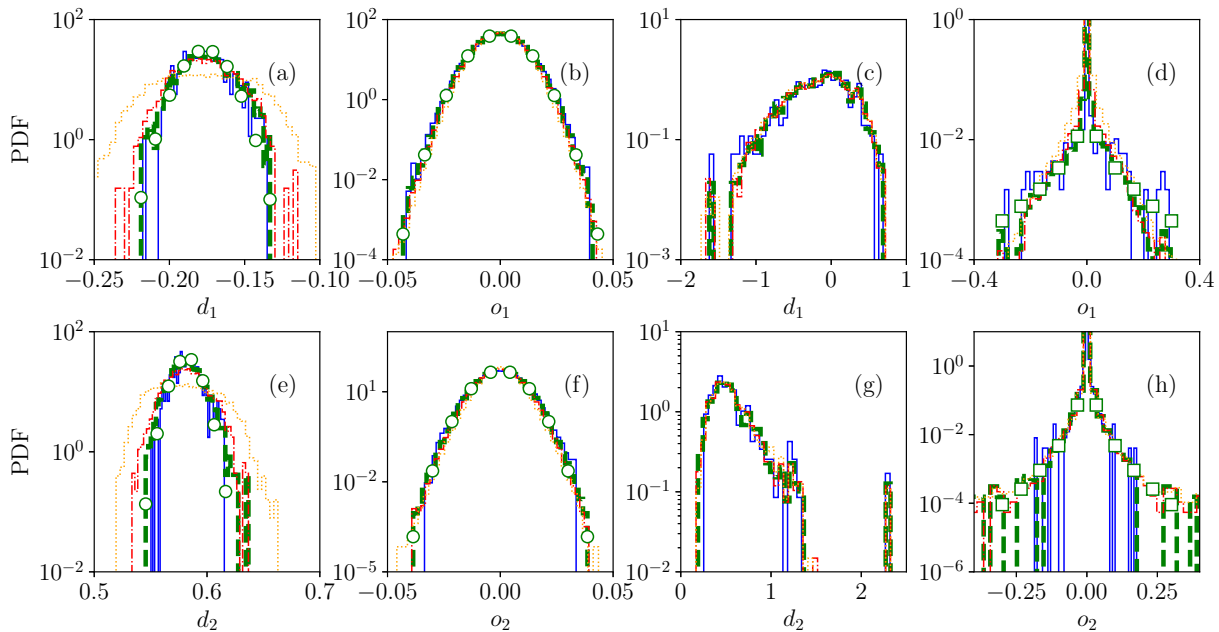}
\caption{Histograms of the matrix elements in the diagonal blocks
  $\tilde{O}^{(0,0)}(\beta=0)$ for $\hat{O}_1$~(upper row)
  and $\tilde{O}^{(0,0)}(\beta=0.2)$ for $\hat{O}_2$~(lower row). 
(a), (b), (e), and (f) are for the nonintegrable case with $\lambda=1$, and
the others for the integrable case with $\lambda=0$.
The energy shell widths are $\Delta_E = 0.1$~(blue solid),
$0.3$~(green dashed), $0.5$~(red dashed-dotted), and $1.0$~(orange dotted).
Diagonal and off-diagonal matrix elements of $\hat{O}_i$ are denoted as $d_i$ and
$o_i$, respectively. The numerical histograms are compared with the Gaussian
distribution functions~(circular symbols) and the stretched exponential
functions~(square symbols).}
  \label{fig:2}
\end{figure*}
\subsection{Diagonal blocks}\label{subsec:dia}
We focus on a diagonal block $\tilde{O}^{(0,0)}(\beta)$ which is characterized with
$E_{\gamma\alpha} \simeq \bar{E}(\beta)$ and $\omega_{\gamma\alpha}\simeq
0$.
A diagonal block includes both diagonal and off-diagonal matrix elements. 
Thus, one can compare the distributions of both quantities.
Figure~\ref{fig:2} presents the histograms obtained at 
$\beta=0.0$ for $\hat{O}_1$ and $0.2$ for $\hat{O_2}$. 

Before proceeding, we remark on the effect of the energy shell size
$\Delta_E$~(see also discussions in Ref.~\cite{Richter:2020bx}). 
In Fig.~\ref{fig:2}, we compare the histograms obtained with
$\Delta_E=0.1,~ 0.3,~0.5$, and $1.0$.
There is a tradeoff between small and large $\Delta_E$. For small
$\Delta_E$, one can study the intrinsic statistical property at a given
energy scale. However, statistics becomes worse because an energy
shell includes fewer eigenstates. 
As $\Delta_E$ increases, statistics becomes better but 
a systematic correction arises.
For instance, in Fig.~\ref{fig:2} (a) and (e), the histogram broadens as 
$\Delta_E$ increases. 
A diagonal element $O_{\alpha\alpha}$ is an energy 
eigenstate expectation value. With finite $\Delta_E$, the histogram is given
by the superposition of the intrinsic distribution functions at different
energy values. The extrinsic fluctuation due to the energy dispersion results 
in the broadening. 
{In order to suppress the extrinsic fluctuation, 
$\Delta_E$ should be smaller than the energy dispersion 
$\delta E = O(L^{d/2})$ with dimensionality $d$ of 
the thermal equilibrium state.}
In this paper, we will use the intermediate value of $\Delta_E = 0.3$ unless
stated otherwise. 
The block submatrix $\tilde{O}^{(0,0)}(\beta)$ is of size 
$\mathcal{N}\times\mathcal{N}$ where $\mathcal{N} = 1278~(1062)$ for $\beta =
0.0~(0.2)$ when $\lambda = 1$ and $\mathcal{N} = 916~(741)$ for $\beta =
0.0~(0.2)$ when $\lambda=0$.

The diagonal and off-diagonal elements of both observables follow the Gaussian distribution 
in the nonintegrable case. In Figs.~\ref{fig:2}(b) and (f), we compare the numerical histogram of the 
off-diagonal elements with the symmetric Gaussian function,
represented with circular symbols, of the same variance. They are in perfect
agreement. 
The ETH predicts that the variance of the diagonal elements
is twice that of the off-diagonal elements~\cite{Mondaini:2017jg,Jansen:2019hu}. 
In Figs.~\ref{fig:2}(a) and (e), we compare the numerical histogram
of the diagonal elements with the Gaussian function whose 
variance is set to a double of the variance of the off-diagonal elements.
The center of the Gaussian is shifted to the mean value of the diagonal elements.
The perfect agreement confirms the ETH ansatz in Eq.~\eqref{ETH_ansatz} for
$\omega_{\gamma\alpha}\simeq 0$.

In the integrable case, the numerical data are not compatible with the ETH ansatz.
The histograms of diagonal elements, shown in
Figs.~\ref{fig:2}(c) and (g), and of off-diagonal elements, 
shown in Figs.~\ref{fig:2}(d) and (h), do not have the Gaussian shape.
The histogram of the off-diagonal elements is fitted well with a stretched
exponential function $f_s(x) = p e^{-q|x|^\theta}$, which is plotted with
square symbols in Fig.~\ref{fig:2}(d) and (h). The exponent takes 
a value $\theta \simeq 0.47$ in Fig.~\ref{fig:2}(d) and $\theta\simeq 0.34$ 
in Fig.~\ref{fig:2}(h). Our numerical results suggest that the exponent 
may vary with $\beta$. 
Statistics of the off-diagonal elements of $\hat{O}_1$ is also 
investigated in Ref.~\cite{LeBlond:2019bv} at different values of $\Delta$.
It is reported that they may follow a log-normal distribution. 
A theoretical study is necessary in order to understand the nature of the
offdiagonal elements in the integrable systems, which is beyond the scope of
the current work.

The variance of off-diagonal elements, denoted by $\sigma^2_o[\hat{O}]$ 
with subscript $o$ standing for off-diagonal elements, depends on
the system size $L$. We investigate the size dependence 
at $\beta = 0.0,~ 0.1,~ 0.2$, and $0.3$.
The ETH predicts that 
\begin{equation}
  \sigma^2_{o}[\hat{O}] = e^{-S(\bar{E})} |f_O(\bar{E},\omega = 0)|^2 .
  \label{VarO_nonint}
\end{equation}
The system size dependence comes into play through the entropy function 
$S(\bar{E}(\beta))$. It can be estimated as $S(\bar{E}) = \ln
\mathcal{D}(\bar{E})$ where $\mathcal{D}(\bar{E}) =
\mathcal{N}_0/\Delta_E$ is the density of the states.

\begin{figure}[t]
  \includegraphics*[width=\columnwidth]{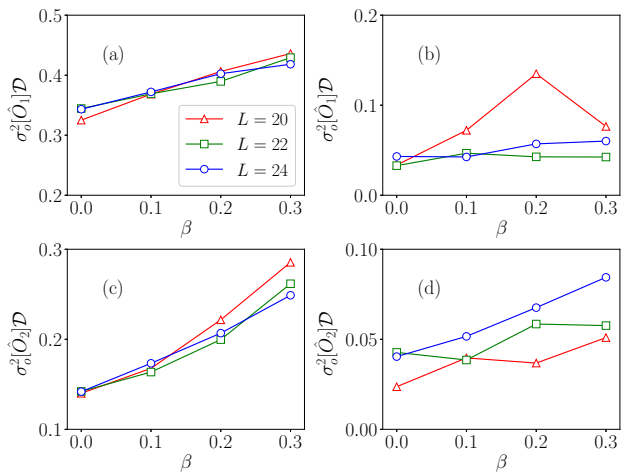}
  \caption{Scaled variance of off-diagonal elements within a diagonal
    block $\tilde{O}^{(0,0)}(\beta)$. The variance is multiplied by the density of states
    $\mathcal{D}(\bar{E}(\beta))$. (a) and (c) correspond
    to the nonintegrable case~($\lambda=1$), while the others corresponds 
    to the integrable case~($\lambda=0$). 
  The system sizes are $L=20$, $22$, and $24$.}
  \label{fig:3}
\end{figure}

The variance multiplied by $\mathcal{D}$ is plotted in Fig.~\ref{fig:3} as a
function of $\beta$. In the nonintegrable case, the curves from different
system sizes $L=20, 22$, and $24$ tend to align along a single curve, which
corresponds to the function $|f_O(\bar{E}(\beta),\omega=0)|^2$ for the
observable $\hat{O_1}$ or $\hat{O_2}$ appearing in Eq.~\eqref{VarO_nonint}. 
This behavior is fully consistent with the prediction of the ETH.
Most of previous works investigated the scaling of the off-diagonal elements 
corresponding to the infinite temperature state~($\beta=0$) where 
$\mathcal{D}$ is proportional to the dimensionality of the total Hilbert 
space~\cite{Mondaini:2017jg,LeBlond:2019bv,Jansen:2019hu}. 
Our work confirms the scaling form of Eq.~\eqref{VarO_nonint} 
at finite temperature states. 

In the integrable system, the scaled variances obtained 
at different system sizes are rather scattered. 
Nevertheless, from the plots in Fig.~\ref{fig:3}(b) 
for $L=22$ and $24$, we expect that the off-diagonal elements of $\hat{O}_1$ 
follow the same scaling form of Eq.~\eqref{VarO_nonint} for large enough
system sizes. The data for $\hat{O}_2$ are more scattered. 
However, further analysis in the following subsection 
supports that the scaling form $\sigma_o^2 \propto 1/\mathcal{D}$ 
is also valid for $\hat{O}_2$. 
Such a scaling has been also reported in the integrable {\it XXZ} model with other
parameter values~\cite{LeBlond:2019bv}.

We remark that the scaling $\sigma^2_o \propto 1/\mathcal{D}$ 
should not be regarded as the evidence for the ETH. We have already
shown that the off-diagonal elements in the
integrable system do not follow the Gaussian distribution. 
What is shown in Fig.~\ref{fig:3} is that the fluctuation
amplitude of the off-diagonal elements scales as $1/\mathcal{D}$ 
in both integrable and nonintegrable systems.

\subsection{Off-diagonal blocks}\label{subsec:off}
\begin{figure}[t]
  \includegraphics*[width=\columnwidth]{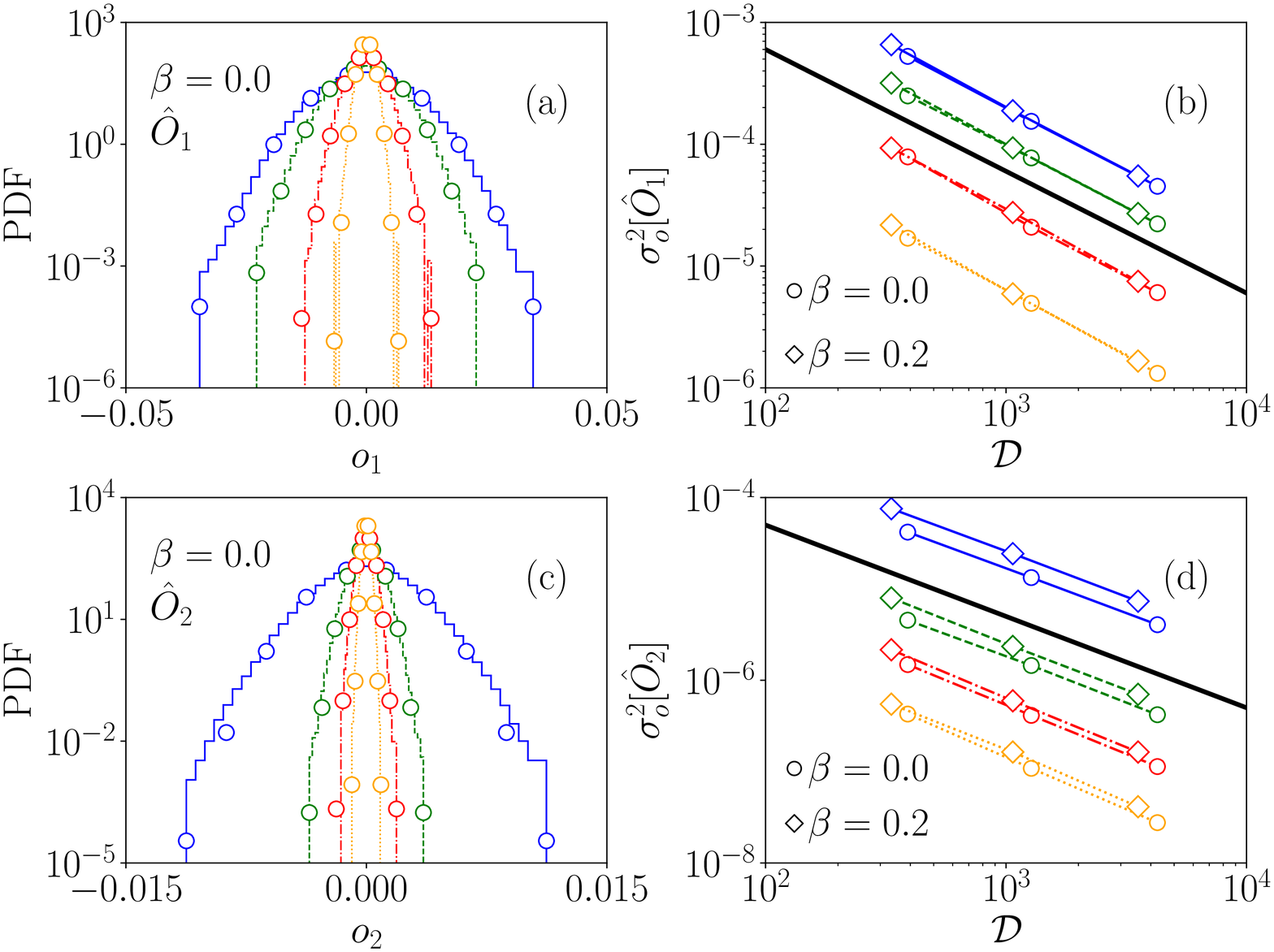}
  \caption{Numerical data for the off-diagonal blocks $\tilde{O}^{(b,-b)}(\beta)$
  with $\omega = 2b \Delta_E$ with 
  $2b = 4$~(blue solid), $8$~(green dashed), $12$~(red dashed
  dotted), and $16$~(orange dotted) in the {\em nonintegrable} case. 
    (a) and (c) show the histograms for $\hat{O}_1$ and $\hat{O}_2$, respectively, 
    at $L=24$ and $\beta=0$.
    Each numerical histogram is compared with the Gaussian function, marked
    with circular symbols, of the same variance.
    The variance is plotted against the density of states
    $\mathcal{D}(\bar{E}(\beta))$ in (b) for $\hat{O}_1$ and (d) for $\hat{O}_2$
    at $\beta=0.0~(\circ)$ and $0.2~(\diamond)$. 
    The thick straight line of slope $-1$ is a guide to the eye. 
    The sizes
    of the block submatrices with $2b=4$ are $(1269\times 1226)$ at
  $\beta=0.0$ and $(1182\times 912)$ at $\beta=0.2$ when $L=24$. 
  The other block submatrices are of similar size.}
  \label{fig:4}
\end{figure}

\begin{figure}[t]
  \includegraphics*[width=\columnwidth]{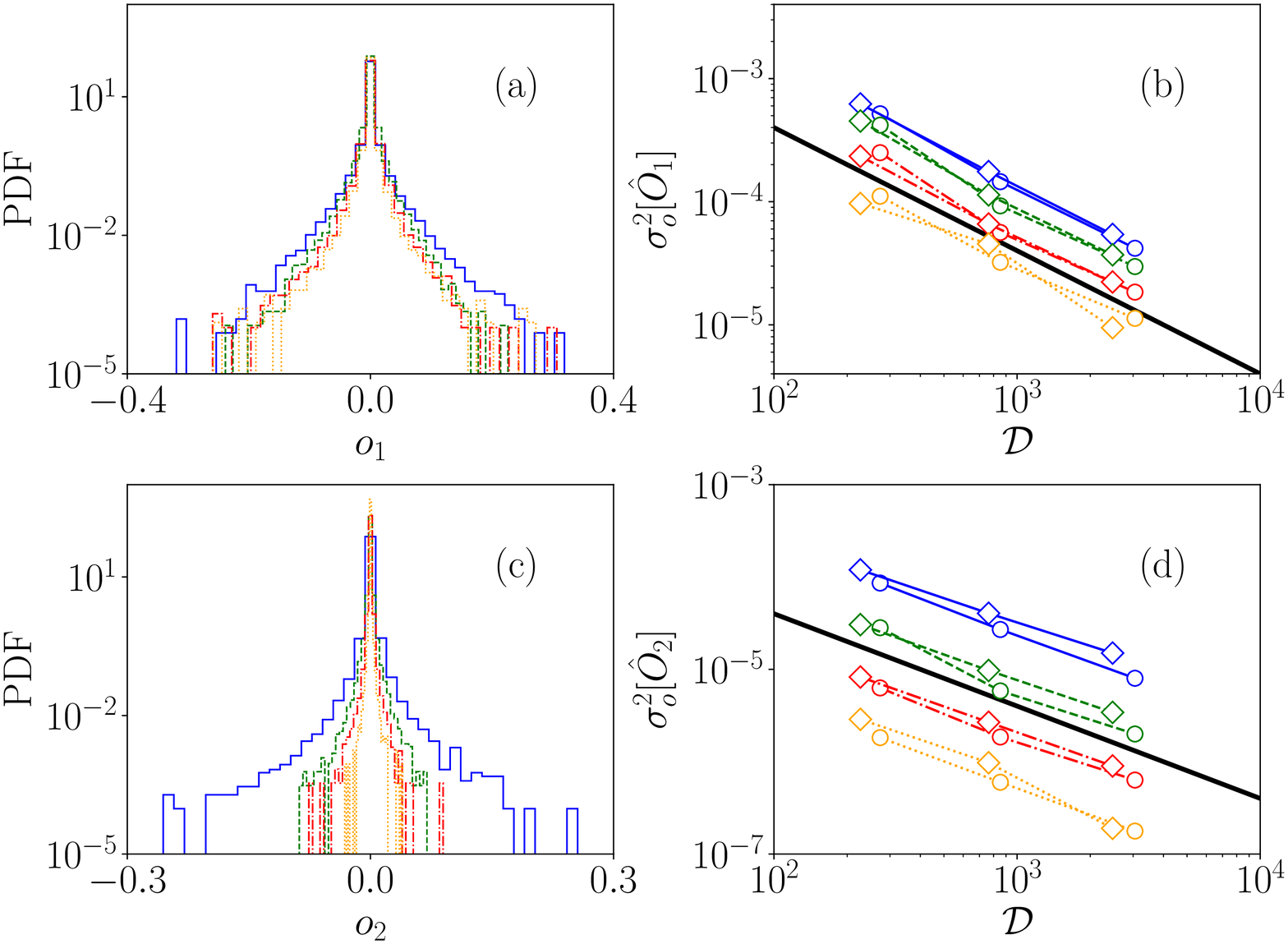}
  \caption{The same plots as in Fig.~\ref{fig:4} for the {\em integrable}
    case. The sizes of the block submatrices with $2b=4$ are $(886\times 895)$
  at $\beta=0.0$ and $(808\times678)$ at $\beta=0.2$ when $L=24$.}
  \label{fig:5}
\end{figure}

We investigate matrix elements of off-diagonal blocks
$\tilde{O}^{(b,a)}(\beta)$ with $b\neq a$. As explained in
Sec.~\ref{sec:model}, a block submatrix $\tilde{O}^{(b,a)}(\beta)$ consists of matrix
elements $\{O_{\gamma\alpha}\}$ with $E_\alpha \simeq \bar{E}(\beta)+a\Delta_E$ and 
$E_\gamma \simeq \bar{E}(\beta)+b\Delta_E$.
Off-diagonal blocks with $b\neq a$ include only off-diagonal elements. 
Specifically, we will consider the block submatrices with
$b=-a~(=1,2, \cdots)$ in which $E_{\gamma\alpha} \simeq \bar{E}(\beta)$ and 
$\omega_{\gamma\alpha} \simeq \omega = 2b\Delta_E$. They are represented by 
the shaded squares in Fig.~\ref{fig:1}.
We have measured the probability distribution and the variance 
at each off-diagonal block. 
The numerical results
at $\beta=0$ and $0.2$ are presented in Fig.~\ref{fig:4} 
for the nonintegrable case and Fig.~\ref{fig:5} for the integrable case. 

{\em Non-integrable case.---} At all values of $\bar{E}(\beta)$ and $\omega$
considered, off-diagonal matrix elements follow the symmetric Gaussian 
distribution.  In Fig.~\ref{fig:4}(a) and (c), we compare the numerical 
histogram obtained at $L=24$ with the symmetric Gaussian function of 
the same variance. The agreement is almost perfect, which 
indicates the validity of the ETH. 

The variance of the elements within $\tilde{O}^{(b,-b)}(\beta)$ 
at different system sizes $L=20$, $22$, and $24$ is plotted as a function 
of the density of states $\mathcal{D}(\bar{E}(\beta))$ 
in Figs.~\ref{fig:4}(b) and (d).
For both observables $\hat{O}_1$ and $\hat{O}_2$, the variance is inversely
proportional to the density of states at all values of $\beta$
and $\omega$. Their slopes correspond to the functions
$|f_O(\bar{E},\omega)|^2$. These numerical results,
together with those in the preceding subsections, strongly support the
ETH ansatz in Eq.~\eqref{ETH_ansatz} at all energy scales.

{\em Integrable case.--} We have performed the same analysis in the
integrable system. The histograms for $\hat{O}_1$ and $\hat{O}_2$ are
presented in Fig.~\ref{fig:5}(a) and (c),  respectively.
As in the case with $\omega=0$~[Figs.~\ref{fig:2}(d) and (h)], 
the distribution functions are non-Gaussian. They decay more slowly 
than an exponential function in the tail.

We also investigate the finite-size scaling behavior of the variance. 
We have measured the variance of the matrix elements in each off-diagonal 
block $\tilde{O}^{(b,-b)}(\beta)$ at different system sizes $L=20$, $22$, and
$24$. They are plotted against the density of states
$\mathcal{D}(\bar{E}(\beta))$ in Figs.~\ref{fig:5}(b) and (d).
The overall behavior is consistent with the scaling
$\sigma^2_{o}(\hat{O}) \propto 1/\mathcal{D}$ for both observables at all
parameter values.
Such a scaling was also reported for the integrable
systems at the center of the energy spectrum~\cite{Beugeling:2015hq,LeBlond:2019bv}.
In comparison to the data shown in Fig.~\ref{fig:4}, the data suffer from stronger 
fluctuations. These behaviors were also observed in the diagonal 
blocks~[see Figs.~\ref{fig:3}(b) and (d)].

In this subsection, we have investigated the statistical 
property of off-diagonal matrix elements $\{O_{\gamma\alpha}\}$ within each block 
characterized with $E_{\gamma\alpha} \simeq \bar{E}(\beta)$
and $\omega_{\gamma\alpha} \simeq \omega \neq 0$. We have confirmed that the
nonintegrable system follows the prediction of the ETH in
Eq.~\eqref{ETH_offd} with Gaussian random variables $R_{\gamma\alpha}$.
In the integrable systems, off-diagonal matrix elements do not follow the
Gaussian distribution. However, the variance is still inversely 
proportional to the density of states.

\section{Eigenstate-to-eigenstate fluctuations}\label{sec:eef}
The ETH ansatz in Eq.~\eqref{ETH_ansatz} is a strong requirement that all
matrix elements involving nearby energy eigenstates should be 
statistically equivalent.
The equivalence for the diagonal elements has been tested by
performing the finite size scaling analysis of $\delta O_\alpha =
|O_{(\alpha+1)(\alpha+1)}-O_{\alpha\alpha}|$~\cite{Kim:2014kw,Mondaini:2016dn,Jansen:2019hu} and of the
deviation of eigenstate expectation values from the microcanonical ensemble
average~\cite{Yoshizawa:2018js}. 
However, off-diagonal elements have been studied only at the {\em coarse-grained
level}. The eigenstate-to-eigenstate fluctuations for off-diagonal
elements have not been studied both for the nonintegrable and the integrable
systems, which will be addressed in this section.

\begin{figure}[t]
  \includegraphics*[width=\columnwidth]{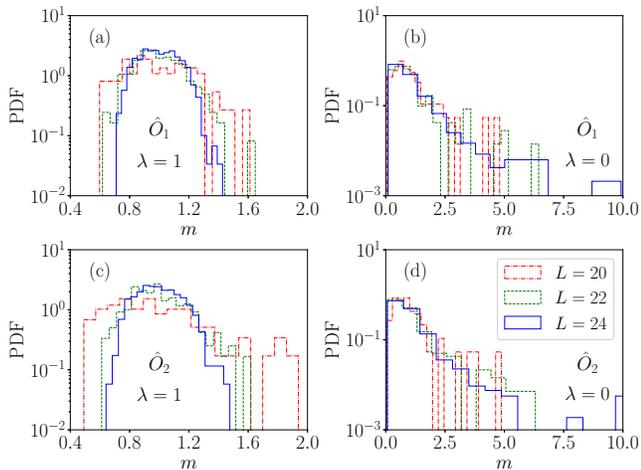}
  \caption{Histogram of the normalized columnar second
    moments, $m=M_2/\langle M_2\rangle$,
    of off-diagonal elements within a block submatrix 
    $\tilde{O}^{(b,-b)}(\beta)$ with $\beta=0.0$ and $b=8$. 
When $L=24$, the sizes of the block submatrices are
$(921\times829)$ in the nonintegrable case and $(712\times 776)$ in the
integrable case.}
  \label{fig:6}
\end{figure}

We consider a block submatrix $\tilde{O}^{(b,-b)}(\beta)$ characterized with
$E_{\gamma\alpha} \simeq \bar{E}(\beta)$ and $\omega_{\gamma\alpha} \simeq
\omega = 2b \Delta_E$. 
A column corresponding to an energy eigenstate $|\alpha\rangle \in
\mathcal{S}_{-b}$ is a set of $O_{\gamma\alpha}$'s where $|\gamma\rangle\in
\mathcal{S}_{b}$~(see Fig.~\ref{fig:1}).
Instead of measuring the moment of all elements in a block, 
we measure the second moment $M_2$ of the elements in a column
separately. Then, we can quantify the eigenstate-to-eigenstate fluctuations 
from the distribution of the columnar second moments.

In Fig.~\ref{fig:6}, we present the histogram of the normalized columnar 
second moments $m \equiv M_2/ \langle M_2\rangle$ where 
$\langle M_2\rangle$ is the mean value of the second moments of all columns
within a block submatrix. In the nonintegrable
case with $\lambda=1$, shown in Figs.~\ref{fig:6}(a) and (c), the distribution 
functions are peaked around $m \simeq 1.0$. Furthermore,
the peak becomes narrower as the system size $L$ increases.

The integrable system~($\lambda=0$) exhibits distinct behaviors. 
The distribution functions shown in Figs.~\ref{fig:6}(b) and (d)
have a long tail. Furthermore, we cannot find any signature suggesting that 
the width of the distribution may decrease with the system size.

\begin{figure}[t]
  \includegraphics[width=\columnwidth]{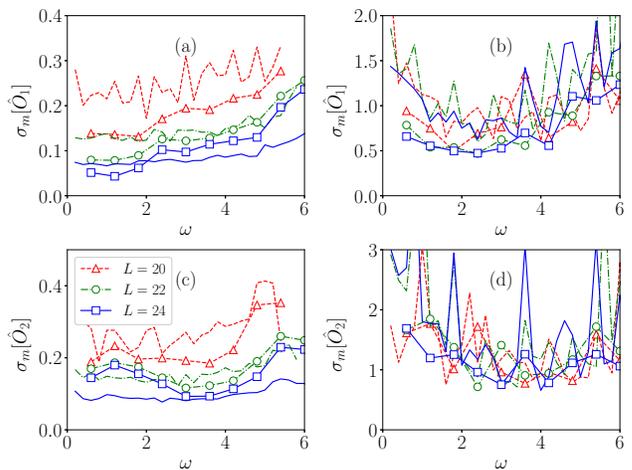}
  \caption{Standard deviation $\sigma_m[\hat{O}_i]$ 
    of the normalized columnar second moments of off-diagonal elements 
    in $\tilde{O}_{b,-b}(\beta=0)$. The data are plotted
    as a function of $\omega = 2b \Delta_E $. Lines with symbols
  represent the data with $\Delta_E = 0.3$, while lines without symbols
represent the data with $\Delta_E=0.1$. The panels (a) and (c) are for the
nonintegrable system, and (b) and (d) are for the integrable system.}
  \label{fig:7}
\end{figure}

We quantify the eigenstate-to-eigenstate fluctuation with the standard
deviation of $m$.
The numerical data are plotted in Fig.~\ref{fig:7}. 
In the nonintegrable case~($\lambda=1$), the standard deviation decreases as
the system size increases at all values of $\omega = 2b\Delta_E$. On the other hand, in
the integrable case~($\lambda=1$), the standard deviation tends to converge
to nonzero values. 

We remark that there are two sources of the fluctuations leading to nonzero
values of $\sigma_m$.
First of all, any intrinsic eigenstate-to-eigenstate fluctuations, which we
are interested in, are responsible for the variation of $m$.
In addition, extrinsic fluctuations due to the energy
dispersion of the order of $\Delta_E$ among eigenstates also contribute to the fluctuations.
In order to reduce the effect of the extrinsic fluctuations, 
we have also measured the standard deviation of $m$ using the smaller values 
of the energy shell width $\Delta_E = 0.1$. 
The two data sets from $\Delta_E=0.1$ and $0.3$ are
compared in Fig.~\ref{fig:7}, which gives a hint on the nature of the intrinsic fluctuations. 
In the nonintegrable case, the relative standard deviation 
decays more rapidly with the system size at $\Delta_E=0.1$. 
In the integrable case, the
relative fluctuation becomes even stronger at $\Delta_E=0.1$.

Based on the numerical results, we conclude that the intrinsic
eigenstate-to-eigenstate fluctuations vanish in the nonintegrable system 
while they remain finite in the integrable system in the infinite size limit.
Our result provides a strong support for the ETH ansatz for off-diagonal
elements in the nonintegrable systems. 
It also reveals that the eigenstate-to-eigenstate fluctuations as well as the
non-Gaussian distribution functions invalidate the ETH for off-diagonal
elements in the integrable systems.

\section{Summary and Discussions}\label{sec:summary}
In summary, we have performed a thorough numerical analysis on the statistical
property of matrix elements $\{O_{\gamma\alpha}\}$ of observables in the
energy eigenstate basis of the integrable and the nonintegrable spin-$1/2$
{\it XXZ}
chains. Using the block submatrix, we
have investigated the statistical property in the energy-resolved way. 
Our study confirms that the ETH ansatz characterizes statistics of matrix
elements in the nonintegrable model at all energy scales. 
In various regions with different
values of $E_{\gamma\alpha} \simeq \bar{E}$ and $\omega_{\gamma\alpha}
\simeq \omega$, the distribution
functions are shown to follow the prediction of the ETH.

Statistics in the integrable system is subtle. Both the diagonal and
off-diagonal elements do not follow the Gaussian distribution of the ETH.
On the other hand, the variance of off-diagonal elements within blocks 
seems to be well-defined and inversely proportional to the density 
of states.
However, we discover that the eigenstate-to-eigenstate fluctuations are relevant in the
integrable system. 
Figures~\ref{fig:6} and
\ref{fig:7} show that the eigenstate-to-eigenstate fluctuations are
of the same order of magnitude as the overall fluctuations, which disproves a
scaling form such as in Eq.~\eqref{VarO_nonint}. 

In the nonintegrable system, the eigenstate-to-eigenstate 
fluctuation vanishes in the large system size limit. It implies 
that the nonintegrable system has the self-averaging property that
nearby energy eigenstates are statistically equivalent to each
others. It provides a strong justification of the ETH.

The eigenstate-to-eigenstate fluctuation is important to understand 
the origin for the breakdown of the FDT in the integrable system.
When a system is prepared initially at an energy eigenstate $\ket{\alpha}$, 
a dynamic correlation function $\bar{S}(\omega) = \int_{-\infty}^\infty dt
\langle \hat{O}(t)\hat{O}(0)\rangle e^{i \omega t}$
of an observable $\hat{O}$ in the frequency domain is given 
by~\footnote{In general, the FDT is formulated for two different
operators. In this work, it suffices to consider the autocorrelation
function.}
\begin{equation}
  \bar{S}(\omega) = 2\pi \sum_{\gamma\neq \alpha} O_{\alpha \gamma}
  O_{\gamma\alpha} \delta(\omega-\omega_{\gamma\alpha}) .
  \label{corr}
\end{equation}
In the Gibbs state, 
the correlation function and the response function are related 
through the FDT~\cite{Mazenko:2006vd}. 
Recently, it is shown that the FDT is also valid in an energy eigenstate 
for the isolated quantum systems obeying the 
ETH~\cite{DAlessio:2016gr,Noh:2020fs,Schuckert:2020bu}. 
The derivation relies on the properties that the second moment of
$O_{\gamma\alpha}$'s is a smooth function of $E_{\gamma\alpha}$ and
$\omega_{\gamma\alpha}$ and is inversely proportional to the
density of states $\mathcal{D}(E_{\gamma\alpha})$.
Interestingly, the numerical results of Sec.~\ref{subsec:off} and 
the literatures, e.g., Ref.~\cite{LeBlond:2019bv}, 
show that the integrable systems have the same properties at the
coarse-grained level.
It is puzzling because the FDT is violated in the integrable 
systems~\cite{Foini:2011kw,Foini:2012gq}. This puzzle is resolved by
the eigenstate-to-eigenstate fluctuations.

The FDT can be casted in the form of the Kubo-Martin-Schwinger~(KMS) 
relation~\cite{Mazenko:2006vd,Haag:1967sg} 
\begin{equation}
  \bar{S}(\omega) = e^{\beta\omega} \bar{S}(-\omega) .
  \label{KMS}
\end{equation}
It constrains that $\frac{1}{\omega}\ln[\bar{S}(\omega)/\bar{S}(-\omega)]$
should depend only on the inverse temperature $\beta$. It should be independent of 
$\omega$ and the eigenstate quantum number $\alpha$.
The correlation function $\bar{S}(\omega)$ in Eq.~\eqref{corr} involves a 
columnar second moment. In the presence of the eigenstate-to-eigenstate
fluctuations, the columnar second moment has an explicit eigenstate dependence, which
results in $\omega$ and $\alpha$ dependence of 
$\frac{1}{\omega} \ln[\bar{S}(\omega) / \bar{S}(-\omega)]$  in general.
It explains the reason why the FDT is violated in the integrable systems.

The eigenstate-to-eigenstate fluctuations presented in Fig.~\ref{fig:7} depend
on the choice of $\Delta_E$. 
In order to reduce the effect of the extrinsic fluctuations and achieve a good statistics, 
one need to choose a smaller value of $\Delta_E$ at larger system sizes. We
leave a more quantitative study on the finite size scaling property at larger
systems as a future work.

\begin{acknowledgments}
  This work is supported by the National Research Foundation of Korea~(NRF)
  grant funded by the Korea government~(MSIP)~(Grant No. 2019R1A2C1009628).
\end{acknowledgments}

\bibliography{paper}

\end{document}